\documentclass[12pt,preprint]{aastex}
\usepackage{amsmath}
\shorttitle{The blazar sequence and inverse Compton models}
\shortauthors{Chen \& Bai}

\slugcomment{Printed at \today}
\slugcomment{Accepted by ApJ}

\begin{document}

\title{Implications on the blazar sequence and inverse Compton models from \emph{Fermi} bright blazars}

\author{Liang Chen\altaffilmark{1,\,2,\,3,\,4\dag} and J.M. Bai\altaffilmark{1,\,3}}

\altaffiltext{1}{National Astronomical Observatories/Yunnan Observatory, Chinese Academy of Sciences, Kunming, 650011, China}
\altaffiltext{2}{Key Laboratory for Research in Galaxies and Cosmology, Shanghai Astronomical Observatory, Chinese Academy of Sciences, 80 Nandan Road, Shanghai, 200030, China}
\altaffiltext{3}{Key Laboratory for the Structure and Evolution of Celestial Objects, Chinese Academy of Sciences, Kunming, 650011, China}
\altaffiltext{4}{The Graduate School of Chinese Academy of Sciences}
\altaffiltext{\dag}{E-mail: chenliangew@hotmail.com; baijinming@ynao.ac.cn}

\def\beq{\begin{equation}}
\def\eeq{\end{equation}}
\def\bar{\begin{eqnarray}}
\def\ear{\end{eqnarray}}
\def\tr{\tau_{rad}\,}
\def\trT{\tau_{rad,\,T}\,}
\def\trKN{\tau_{rad,\,KN}\,}
\def\tesc{\tau_{esc}\,}
\def\tac{\tau_{acc}\,}
\def\tpp{\tau_{pp}\,}
\def\epm{$e^\pm$\,\,}

\begin{abstract}

In this paper, we use the quasi-simultaneous spectra of \emph{Fermi} bright blazars and \emph{Fermi} detected narrow line Seyfert 1
(NLS1) to study the blazar sequence and inverse Compton (IC) models. \textbf{I.} The synchrotron peak luminosities ($L_{s}$) significantly
inverse correlate with the synchrotron peak frequencies ($\nu_{s}$), $L_{s}\propto\nu_{s}^{-0.44\pm0.11}$, which is consistent with the blazar sequence. In addition to the correlation, there are some blazars showing low $\nu_{s}$ and low $L_{s}$. To study the
relation between these low $\nu_{s}$ low $L_{s}$ blazars and the blazar sequence, we present correlations of the parameter
$L_{s}\nu_{s}^{1/4}$ with the ratio of Compton to synchrotron peak frequencies ($r_{Cs}\equiv\nu_{C}/\nu_{s}$) and with the ratio of Compton to synchrotron luminosities ($CD\equiv L_{C}/L_{s}$). The results indicate that both correlations are significant with a Pearson's probability for null correlation of $p=0.0218$ and $p=0.0286$ respectively. This does not support the idea that the low $\nu_{s}$ low $L_{s}$ blazars are sources with less beaming. Another possibility, as suggested by Ghisellini \& Tavecchio, is that these blazars have relative lower black hole masses. To test this, we collect the black hole masses of 30 blazars from archives, and find that the hole mass correlates with the parameter $L_{s}\nu_{s}^{0.44}$ ($p=0.0344$). Therefore, the black hole masses of low $\nu_{s}$ low $L_{s}$ blazars are statistically small. The NLS1s are thought to have lower black hole masses. We find that the four NLS1s detected by \emph{Fermi} have low $\nu_{s}$ and low $L_{s}$. This supports the above result. \textbf{II.} The ratio $r_{Cs}$ correlates with $CD$ significantly ($p=0.00375$). The external Compton (EC) model can naturally explain this correlation, while synchrotron self Compton (SSC) model can not. This agrees with the findings of many authors that the EC process dominates the gamma-ray emission of Flat Spectrum Radio Quasars.

\end{abstract}


\section{Introduction}

Blazars are the most extreme active galactic nuclei (AGNs). Their broadband emissions, from radio through $\gamma$-ray, are dominated by nonthermal emissions produced by relativistic plasma jet aligned the line of sight \citep{1978bllo.conf..328B}. Their spectra energy distribution (SED) show two broad components in $\log\nu-\log\nu L_{\nu}$ diagram. The lower component peaks at infrared (IR) to X-ray bands, which is believed to be the synchrotron emissions of relativistic electrons within jet. The higher component peaks at $\gamma$-ray band, which is thought to be the inverse Compton (IC) emissions of the same electron population. Models are classified according to different origins of the IC seed photons, synchrotron-self Compton \citep[SSC, seed photons from the synchrotron radiation, see][]{1981ApJ...243..700K, 1985ApJ...298..114M, 1989ApJ...340..181G, 1992ApJ...397L...5M, 1985ApJ...298..128B} and external Compton \citep[EC, seed photons from external region, see][]{1992A&A...256L..27D, 1993ApJ...416..458D, 1995ApJ...441...79B, 1994ApJ...421..153S, 2000ApJ...545..107B, 2002ApJ...577...78S}. Blazars are often divided into two subclasses of BL Lacertae objects (BL Lacs) and flat spectrum radio quasars (FSRQs). FSRQs have strong emission lines, while BL Lacs have only very weak or lack the emission lines \citep[equivalent width $<5${\AA}, e.g.,][]{1997A&A...325..109S}.

\citet{1998MNRAS.299..433F} presented a unifying view of the SEDs of blazars, in which both the synchrotron peak luminosity (hereafter $L_{s}\equiv(\nu L_{\nu})_{s}^{p}$) and the Compton dominance (the ratio between Compton and synchrotron luminosities, $CD\equiv L_{C}/L_{s}$) decrease with increasing the synchrotron peak frequency (hereafter $\nu_{s}$). \citet{1998MNRAS.301..451G} modeled the broadband SEDs of 51 $\gamma$-ray loud blazars, and showed that in powerful blazars the radiative energy density is large. The effective IC cooling yields lower electron energy and larger $CD$. The lower energy electron emits at lower frequency. An inverse correlation between $\gamma_{p}$ and $U_{tot}'$ is further derived. Where $\gamma_{p}$ is the electron energy emitting at the synchrotron peak, and $U_{tot}'$ is the summation of the magnetic and radiative energy densities within the Thomson regime. In the following works \citep[e.g.,][]{2002A&A...386..833G, 2009MNRAS.399.2041G, 2010MNRAS.402..497G, 2008MNRAS.385..283C}, the $\gamma_{p}$-$U_{tot}'$ inverse correlation is confirmed. People often call $\nu_{s}-L_{s}$ and/or $\gamma_{p}-U_{tot}'$ the blazar sequence. Large number blazars are detected by \emph{Fermi}/LAT, which are compiled as the LAT Bright AGN Sample \citep[LBAS,][]{2009ApJ...700..597A} and the First LAT AGN Catalog \citep[1LAC,][]{2010ApJ...715..429A}. Both LBAS and 1LAC show the correlations between the $\gamma$-ray luminosity ($L_{\gamma}$) and photon indices ($\Gamma_{\gamma}$). The photon indices correlate with peak frequencies and the $\gamma$-ray luminosity can represent the peak luminosity roughly \citep[see,][]{2010ApJ...715..429A, 2009arXiv0912.2040A}. Therefore, it seems to support the balzar sequence.

Many contrary arguments are also reported \citep{2001ASPC..227..116G, 2004MNRAS.348..937C, 2005MNRAS.356..225A, 2006A&A...445..441N, 2007Ap&SS.309...63P}. They mainly focus on three points. Firstly, many low peak frequency low power blazars are found. This causes no significant correlation between $\log\nu_{s}$ and $\log L_{s}$. Secondly, several high peak frequency FSRQs are reported, in contrast with the correlation mentioned above. The SED properties of these sources are mainly determined from composite spectral indices\footnote{The composite spectra index, $\alpha_{12}$, is usually used to measure the overall trend of the broadband spectra when lacks more detailed spectra information. It is defined as $f_{\nu_{1}}/f_{\nu_{2}}=(\nu_{1}/\nu_{2})^{-\alpha_{12}}$, where $f_{\nu_{1,2}}$ are the flux densities at frequencies $\nu_{1,2}$ \citep{1985ApJ...298..630L}.} rather than from broad band SEDs. It causes the uncertainties of the result \citep[see][]{2007Ap&SS.309...63P}. \citet{2008IJMPD..17.1457M} re-studied these FSRQs and found that they do follow the $\log\nu_{s}-\log L_{s}$ sequence. Thirdly, the blazar sequence predicts that blazars with higher peak frequency (mainly BL Lacs) should be more numerous than blazars with lower peak frequency. However, this prediction has not been proved. As indicated by \citet{2008MNRAS.387.1669G}, the reason may be that the samples considered are flux limited, introducing a bias against low luminosity/high peak frequency blazars. \emph{Fermi}/LAT sensitivity is better than that of \emph{EGRET}, especially for harder spectra \citep{2009ApJ...700..597A, 2010ApJ...715..429A}. Very recently, an interesting finding is that the fraction of the BL Lacs in $\gamma$-ray blazars increases from \emph{EGRET} to \emph{Fermi}/LAT \citep[see][]{1999ApJS..123...79H, 2009ApJ...700..597A, 2010ApJ...715..429A}. In 1LAC \citep{2010ApJ...715..429A}, the number of BL Lacs is even larger than the number of FSRQs.

The first objection mentioned above is the strongest evidence against the blazar sequence. \citet{2008MNRAS.387.1669G} presented that there are two possibilities account for it. The first explanation is that those low $\nu_{s}$ low $L_{s}$ sources may be misaligned. The weak beaming effect would shift blazars to low peak frequency low observed luminosity. The second explanation is that sources with low luminosity and low $\nu_{s}$ may be associated with black holes of smaller mass. The jets of these sources will dissipate energy within the broad line region (BLR). The electrons then cool efficiently, and emit at low frequency \citep{2008MNRAS.387.1669G}.

The blazar sequence constrains our understanding on jet physics. It relates to jet energy dissipation, particle acceleration, the emission region properties and environments, etc. In this paper, we collect the black hole masses and use the quasi-simultaneous broadband SEDs of \emph{Fermi} bright blazars \citep{2009ApJ...700..597A, 2009arXiv0912.2040A} and the SEDs of four \emph{Fermi} detected narrow line Seyfert 1 \citep[NLS1,][]{2009ApJ...707L.142A} to study the blazar sequence. In addition, we also study the EC/SSC models.

In section 2, we discuss the sample. Section 3 discusses the relations between our result and the blazar sequence. Section 4 discusses the inverse Compton (IC) models. We summarize and discuss our findings in Section 5. The cosmology with H$_{0}=70$ km s$^{-1}$Mpc$^{-1}$, $\Omega_{m}=0.3$ and $\Omega_{\Lambda}=0.7$ is adopted throughout the paper.

\section{The Sample}

The first three months operation of \emph{Fermi}-LAT reveals more than 100 blazars ($>10\sigma$), and named as the \emph{Fermi} LAT Bright AGN Sample \citep[LBAS,][]{2009ApJ...700..597A}. \citet{2009arXiv0912.2040A} presented quasi-simultaneous SEDs for 48 LBAS blazars, whose data were collected from radio through $\gamma$-ray within those three months operation. The IC and the synchrotron peak frequencies/fluxes are estimated by fitting the two components with a third degree polynomial of $\nu F_{\nu}=a\cdot\nu^{3}+b\cdot\nu^{2}+c\cdot\nu+d$. There are 43 of these 48 sources having measured redshifts. The peak luminosities and frequencies (in AGN frame) of these blazars can be calculated through $L_{s,C}=4\pi d_{L}^{2}\left(\nu f_{\nu}\right)_{s,C}^{p}$ and $\nu_{s,C}=\left(1+z\right)\nu_{s,C}^{p\_obs}$, where $d_{L}$ is the luminosity distance. The results are listed in table \ref{blazar}. Column (1) provides the LAT name of the source. Columns (2) and (3) indicate the synchrotron peak frequency and luminosity. Columns (4) and (5) denote the IC peak frequency and luminosity. The redshift, $\gamma$-ray photon indices $\Gamma_{\gamma}$, $\gamma$-ray luminosity $L_{\gamma}$ and the optical classification are listed in columns (6), (7), (8) and (9), respectively. Columns (10) and (11) are the black hole masses and the references. For columns (7), (8), (10) and (11) see below.

\section{Implications on the Blazar Sequence}

As discussed above, both LBAS and 1LAC show the correlations between $\gamma$-ray photon indices $\Gamma_{\gamma}$ and $\gamma$-ray luminosity $L_{\gamma}$. Because the spectra index correlates with the synchrotron peak frequency \citep[see e.g.,][]{2010ApJ...715..429A}, the correlation between $\Gamma_{\gamma}$ and $L_{\gamma}$ can be thought as evidence to support the blazar sequence (but see discussion below). Here we use the peak frequency directly to test the sequence.

Figure \ref{nu_lum} shows the correlation between the peak frequency ($\nu_{s}$) and luminosity ($L_{s}$). In which, squares are for those 43 sources (the opened circles are NLS1s, see below). It can be seen that the luminosity statistically decreases with increasing the peak frequency. The solid line presents the best fitting (excluding the NLS1s), which gives $L_{s}\propto\nu_{s}^{-0.44\pm0.11}$ and Pearson's $prob$-value (the significance level at which the null hypothesis of zero correlation is disproved) $p=2.06\times10^{-4}$. This is consistent with those studies using $\Gamma_{\gamma}$ and $L_{\gamma}$ \citep[e.g.,][]{2009MNRAS.396L.105G, 2009ApJ...700..597A, 2010ApJ...715..429A} and supports the blazar sequence. But it also can be seen (see figure \ref{nu_lum}), in addition to statistical inverse correlation, that there presents some sources with low $\nu_{s}$ and low $L_{s}$. This makes the $\log\nu_{s}-\log L_{s}$ plane more like wedge-shape. This result has been presented in previous studies, which yield less significant correlation between $\log\nu_{s}$ and $\log L_{s}$ and taken as opponent evidence to the blazar sequence \citep[e.g.,][]{2001ASPC..227..116G, 2004MNRAS.348..937C, 2005MNRAS.356..225A, 2006A&A...445..441N, 2007Ap&SS.309...63P}.

Additionally, we present the correlation between the Compton dominance ($CD$) and luminosity ($L_{s}$), which gives $p=0.00307$ (see figure \ref{lum_cd}). This result is consistent with another statement of the blazar sequence, which claims inverse correlation between luminosity and the Compton dominance. This is first time using quasi-simultaneous broadband data to confirm the statement. From figures \ref{nu_lum} and \ref{lum_cd}, it is expected that low $\nu_{s}$ low $L_{s}$ sources would have lower $CD$. We plot $\nu_{s}$ vs. $CD$ plane (figure not supplied here), which is also wedge-shape. \citet{2008MNRAS.387.1669G} suggested those low $\nu_{s}$ low $L_{s}$ blazars may be misaligned or have smaller black holes.

If those sources have relative larger viewing angles, they become lower luminosity and lower peak frequency. As we know, the Compton and synchrotron peak frequencies are dependent on the beaming effect with the same way. Therefore, the ratio between Compton to synchrotron peak frequencies $r_{Cs}\equiv\nu_{C}/\nu_{s}$ should be independent on viewing angle. And so does the Compton dominance $CD\equiv L_{C}/L_{s}$. Luminosity is proportional to $\delta^{4}$ and frequency is proportional to $\delta$, where $\delta\equiv1/\left\{\Gamma\left(1-\beta\cos\theta\right)\right\}$ is the beaming factor, $\Gamma=1/\left(1-\beta^{2}\right)$ is the Lorentz factor, $\beta\equiv\upsilon/c$ is the velocity in unit of lightspeed and $\theta$ is the viewing angle. Therefore it is expected that $r_{Cs}$ and $CD$ will be independent on the parameter $L_{s}\nu_{s}^{1/4}$ if the difference really relies on the beaming effect. Hence, we present the correlation between the parameter $L_{s}\nu_{s}^{1/4}$ and $r_{Cs}$ in figure \ref{beaming_mass_r}. Figure \ref{beaming_mass_cd} is the correlation between $L_{s}\nu_{s}^{1/4}$ and $CD$. From figure \ref{beaming_mass_r}, we can see that there is a blazar, 0FGL J1719.3+1746, having extreme ratio $r_{Cs}$ (the triangle at top left corner). From SED of 0FGL J1719.3+1746 \citep[see][]{2009arXiv0912.2040A}, we can see that the IC peak frequency is overestimated. Excluding 0FGL J1719.3+1746, the parameter $L_{s}\nu_{s}^{1/4}$ is correlated with the ratio $r_{Cs}$ although have large scattering ($p=0.0218$). Similar result is derived for $L_{s}\nu_{s}^{1/4}$ vs. $CD$ ($p=0.0286$, see figure \ref{beaming_mass_cd}). This do not support the idea that low $\nu_{s}$ low $L_{s}$ sources are misaligned.

As suggested by \citet{2008MNRAS.387.1669G}, those low $\nu_{s}$ low $L_{s}$ blazars may have smaller black holes \citep{2008MNRAS.387.1669G}, and the jet will dissipate energy within the BLR. This will cause efficient cooling of the electron, and yields low frequency low power \citep[see][]{2008MNRAS.387.1669G}. The low black hole mass also produces the lower Compton dominance \citep[see][]{2008MNRAS.387.1669G}. To check if the black hole masses account for those low $\nu_{s}$ low $L_{s}$ blazars, we collect black hole masses from previous works.

Many authors derived the black hole masses of balzars from different ways \citep[e.g.,][]{2010MNRAS.402..497G, 2002MNRAS.331..111C, 2009RAA.....9.1192C, 2010arXiv1011.5879D, 2003MNRAS.343..505F, 2004ApJ...602..103F, 2003MNRAS.340..632L, 2008MNRAS.385..119W, 2003ApJ...583..134B, 2003ApJ...595..624F, 2001MNRAS.327.1111G, 2006ApJ...637..669L, 2005MNRAS.361..919P, 2004ApJ...615L...9W, 2005ApJ...631..762W, 2002A&A...389..742W, 2004AJ....127...53X, 2005AJ....130.2506X}. Through all papers we know, we collect 30 black hole masses of these 43 blazars. Some blazars were studied by many authors and different hole masses are derived. To reduce the uncertainty, we try to select the hole masses from a unity paper and the uniform method deriving the hole mass.
The result is presented in table \ref{blazar}. Columns (10) and (11) are for black hole masses and the references.

The best fitting of figure \ref{nu_lum} shows $L_{s}\propto\nu_{s}^{-0.44\pm0.11}$. Therefore, the correlation between parameter $L_{s}\nu_{s}^{0.44}$ and hole masses could be used to check if these low $\nu_{s}$ low $L_{s}$ blazars have lower hole masses. Figure \ref{low_mass} presents the result, and the best fitting indicates $p=0.0344$. Despite the scattering, our result supports that low $\nu_{s}$ low $L_{s}$ blazars have smaller black hole \citep[see][]{2008MNRAS.387.1669G}. In order to find more evidences, we use broadband SEDs of 4 radio loud narrow line Seyfert 1 (NLS1) detected by \emph{Fermi}/LAT \citep{2009ApJ...707L.142A} to check the above result. NLS1 is thought to have smaller black hole \citep[e.g.,][and references therein]{2008ApJ...685..801Y}. These 4 radio loud NLS1s are believed to have similar central mechanisms as in blazars \citep[see][]{2009ApJ...699..976A, 2009ApJ...707..727A, 2009ApJ...707L.142A}. Therefore, if our above result is correct, these NLS1s should be in low $\nu_{s}$ and low $L_{s}$ region. We collect the broadband SEDs of these four NLS1s \citep[from \emph{NED}\footnote{http://nedwww.ipac.caltech.edu/} and][]{2009ApJ...707L.142A}. For simplicity, we use two-order polynomial to fit the synchrotron component in $\log\nu-\log\nu L_{\nu}$ diagram. The peak frequency and luminosity are presented in table \ref{NLS1}. We plot this in figure \ref{nu_lum}, which are shown as opened circles. It can be seen that these four sources do have low $\nu_{s}$ and low $L_{s}$ values. This supports our above result.

\section{Implications on Inverse Compton Models}

From discussion in above section, we know that both the ratio $r_{Cs}$ and the Compton dominance $CD$ correlate with the parameter $L_{s}\nu_{s}^{1/4}$. This indicates that $r_{Cs}$ and $CD$ may correlate with each other, although we do not know what the correlation implies. Figure \ref{EC3} shows the plane of $r_{Cs}$ vs. $CD$. The best fitting gives $p=0.00375$ (excluding the blazar 0FGL J1719.3+1746). This is a new result. We will discuss its implications on the emission models (i.e., SSC vs. EC). Of course, no matter what conclusion is derived, it works on statistics. After following discussion, it will be seen that the EC model can predict this correlation naturally, while SSC model can not.

Within the symmetrical sphere model, if an electron population emits the broadband SED of blazar, the synchrotron peak frequency ($\nu_{s}$) corresponds to a peak electron energy \citep[$\gamma_{p}$ in $\gamma-\gamma^{3}N_{\gamma}$ diagram,][]{1998ApJ...509..608T},

\begin{equation}\label{eq_syn_fre}
\nu_{s}=\frac{4}{3}\nu_{L}\gamma_{p}^{2}\delta,
\end{equation}
where $\nu_{L}=eB/(2\pi m_{e}c)$ is the Larmor frequency. If the external radiation is prominent at frequency $\nu_{ext}$, the EC component peaks at \citep[inverse Compton scatter within Thomson regime,][]{1970RvMP...42..237B, 1990MNRAS.245..453C, 1998ApJ...509..608T, 2008MNRAS.387.1669G},

\begin{equation}\label{eq_ec_fre}
\nu_{EC}^{p}=\frac{4}{3}\nu_{ext}\gamma_{p}^{2}\Gamma\delta,
\end{equation}
where $\Gamma$ is the jet Lorentz factor. If there is the EC dominant, the EC and synchrotron luminosities follow \citep{1996MNRAS.280...67G, 1998ApJ...509..608T, 2008MNRAS.387.1669G},

\begin{equation}\label{eq_syn_ec_lum}
\frac{L_{EC}}{L_{sy}}=\frac{U_{ext}'}{U_{B}}\simeq\frac{17}{12}\frac{\Gamma^{2}U_{ext}}{U_{B}},
\end{equation}
where $U_{ext}$ is energy density of external photons in the rest frame of the source, $U_{ext}'\simeq(17/12)\Gamma^{2}U_{ext}$ is that measured in the jet comoving frame, and $U_{B}\equiv B^{2}/8\pi$ is the magnetic field energy density.

Combining equations \ref{eq_syn_fre}-\ref{eq_syn_ec_lum} yields,

\begin{equation}\label{eq_cd_rcs}
\frac{L_{EC}}{L_{sy}}\simeq\frac{17e^{2}}{6\pi m_{e}^{2}c^{2}}\frac{U_{ext}}{\nu_{ext}^{2}}\left(\frac{\nu_{EC}^{p}}{\nu_{s}}\right)^{2}.
\end{equation}
Thus we expect $L_{EC}/L_{sy}\propto\left(\nu_{EC}^{p}/\nu_{s}\right)^{2}$ if the external radiation is constant.

For SSC, the IC emissions rely on the synchrotron emissions. Therefore, the simple relation between $CD$ and $r_{Cs}$ can not be derived.

As suggest by \citet{1998MNRAS.301..451G} \citep[see also][]{1999A&A...341...74H, 2006ApJ...646....8F, 2002A&A...386..833G, 2008MNRAS.385..283C}, the external photons of most blazars are contributed by BLR. And the BLR emissions can be almost uniformly taken as $U_{BLR}\simeq2.65\times10^{-2}{\rm erg\ cm}^{-3}$ and $\nu_{BLR}\simeq2\times10^{15}$Hz \citep[see][]{2008MNRAS.387.1669G}. In this case, $CD=L_{C}/L_{s}\simeq L_{EC}/L_{sy}$ correlates with $r_{Cs}$. The statistical correlation shown in figure \ref{EC3} between $CD$ and $r_{Cs}$ may suggest that most blazars are EC dominant. However this is only qualitative result, because the slope of the best fitting ($s\approx0.4$) is not equal to the predicted slope $s=2$. On the other hand, it is interesting to note that if we use the relation $L_{C}/L_{s}\propto\left(\nu_{C}/\nu_{s}\right)^{2}$ to fit the data, the best fitting $\left(U_{ext}/\nu_{ext}^{2}\right)_{fit}$ does not significantly depart from the BLR value: $\left(U_{ext}/\nu_{ext}^{2}\right)_{fit}\simeq3.2\left(U_{BLR}/\nu_{BLR}^{2}\right)$ (corresponding to the dashed line in figure \ref{EC3}). \citet{2009MNRAS.399.2041G} and \citet{2010MNRAS.402..497G} modeled the SEDs of the \emph{Fermi} bright blazars in detail and suggest that most blazars are EC dominant \citep[see also][]{2009ApJ...704...38S}. Our result is consistent with that.

\section{Discussion}

Because the sample is small, FSRQs and BL Lacs are combined as a uniform class in our study. Although they divide by any criterion \citep[e.g., the Eddington ratio $\dot{m}\sim0.01$, see][and references therein]{2009MNRAS.396L.105G,2009ApJ...694L.107X}, their properties vary continuously. In discussing \emph{Fermi} detected blazars, people sometimes use terms Low Synchrotron Peaked blazars (LSP), Intermediate Synchrotron Peaked blazars (ISP) and High Synchrotron Peaked blazars (HSP) instead of FSRQs and BL Lacs  \citep[e.g.,][]{2009arXiv0912.2040A, 2010ApJ...715..429A}. Throughout this paper we consider them as a single calss. If the sample is enlarged, different subclasses can be separately studied in detail.

\begin{figure}
\epsscale{1}
\plotone{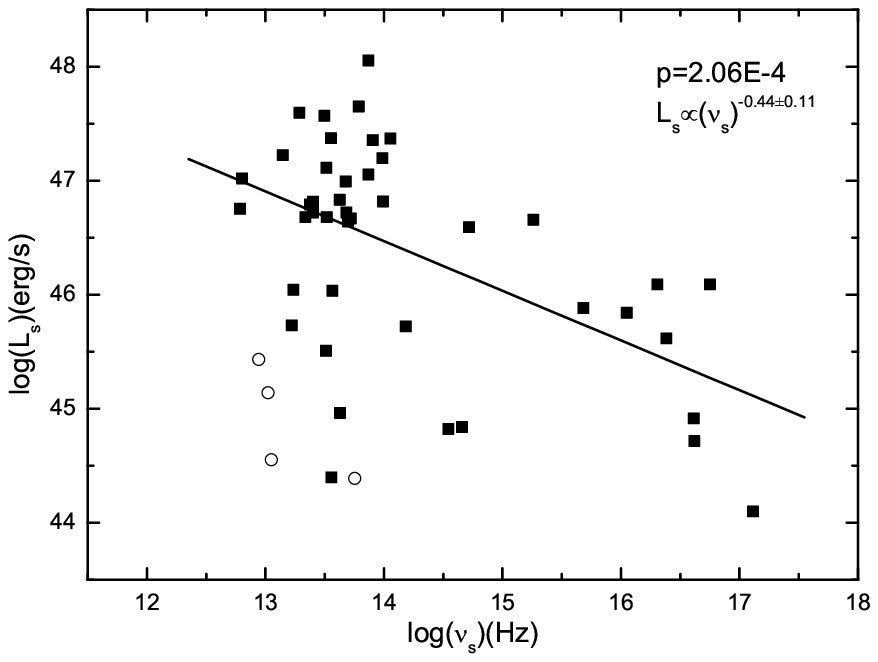}
\caption{The synchrotron peak frequency correlate with the peak luminosity. The squares are balzars \citep{2009ApJ...700..597A, 2009arXiv0912.2040A}. The solid line shows the best fitting with $p=2.06\times10^{-4}$. The opened circles are NLS1s \citep{2009ApJ...707L.142A}.}
\label{nu_lum}
\end{figure}

\begin{figure}
\epsscale{1}
\plotone{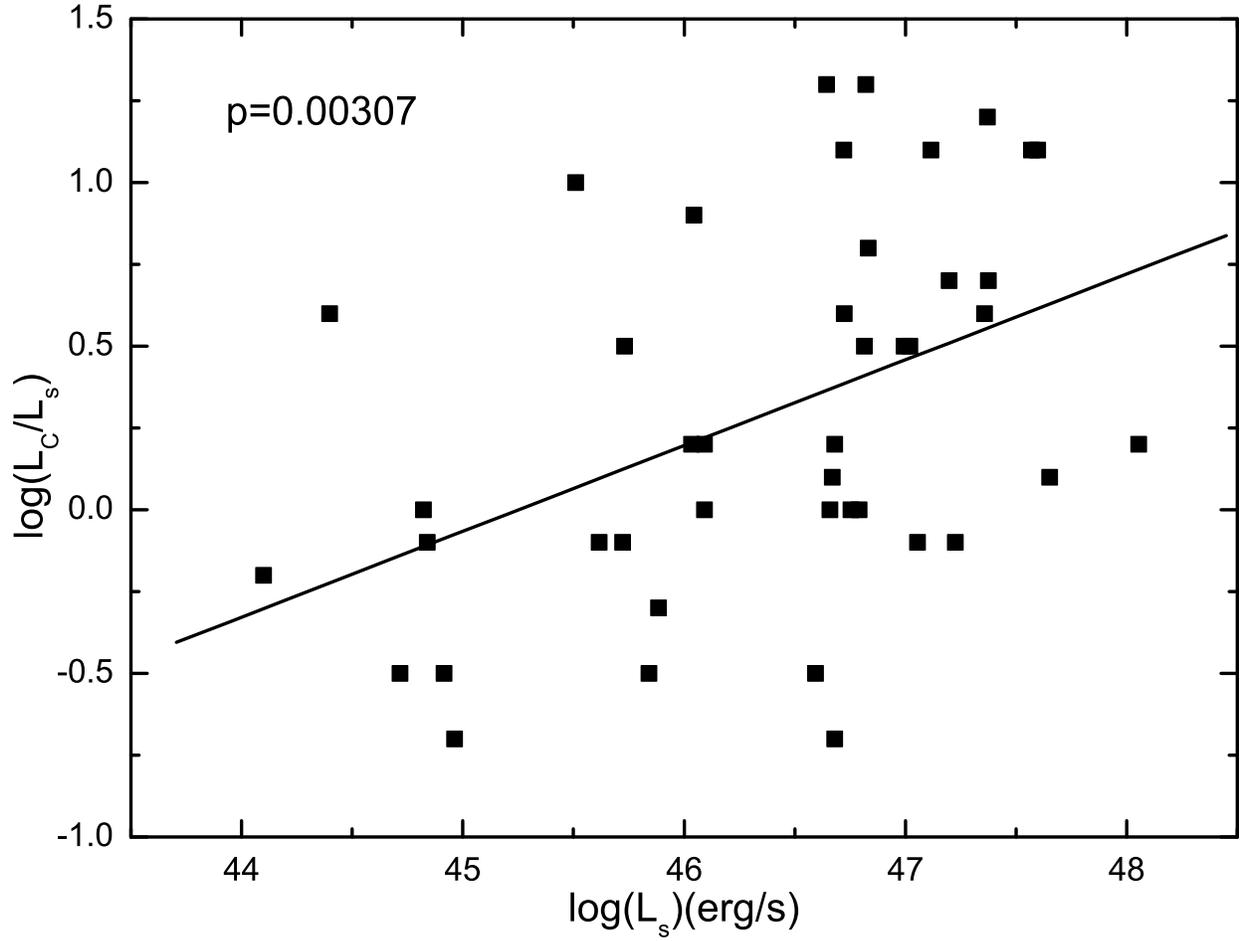}
\caption{The correlation between peak luminosity and the ratio of Compton to synchrotron luminosities $CD\equiv L_{C}/L_{s}$. The solid line shows the best fitting with $p=0.00307$.}
\label{lum_cd}
\end{figure}

Our result of $\log\nu_{s}$ vs. $\log L_{s}$ plane is similar to the result of e.g., \citet{2007Ap&SS.309...63P}. Although the latter study is based on large radio- or X-ray-selected samples, while ours is based on a gamma-ray selected sample, in both of them blazars with low $\nu_{s}$ low $L_{s}$ are presented. In the former study the absence of gamma-ray data does not allow to determine the IC component, therefore the properties of Compton dominance (CD) can not be studied. Their studies and our results indicate that no blazars with high high $\nu_{s}$ high $L_{s}$ have been detected up to now.
\citet{2009MNRAS.396L.105G} \citep[see also][]{2009ApJ...700..597A} studied the \emph{Fermi} bright blazars and showed presence of inverse correlations between $L_{\gamma}$ and $\Gamma_{\gamma}$. As suggested by them, lowering the $\gamma$-ray flux threshold will detect blazars with steep spectral indices and lower luminosities. Here, we notice an interesting thing, which is that if one plots $\log L_{\gamma}$ vs. $\Gamma_{\gamma}$ plane, there is nearly clear inverse correlation \citep[see][]{2009MNRAS.396L.105G, 2009ApJ...700..597A}. While we plot $\log\nu_{s}$ vs. $\log L_{s}$ plane in this paper (see figure \ref{nu_lum}), in addition of inverse correlation, there present some low $\nu_{s}$ low $L_{s}$ blazars. Therefore, when one says the photon index correlates with peak frequency and $\gamma$-ray luminosity correlates with peak luminosity, one should be careful. To check this, we calculate the $\gamma$-ray luminosity of those 43 blazars. The formula we used are similar to that used in \citet{2009MNRAS.396L.105G}. The values are presented in Table \ref{blazar} (see the columns (8) and (9)). We plot $\log L_{\gamma}$ vs. $\Gamma_{\gamma}$ in figure \ref{lum_gamma}. The plane is similar to that in \citet{2009MNRAS.396L.105G}, which shows clear correlation ($p=2.71\times10^{-5}$).

\begin{figure}
\epsscale{1}
\plotone{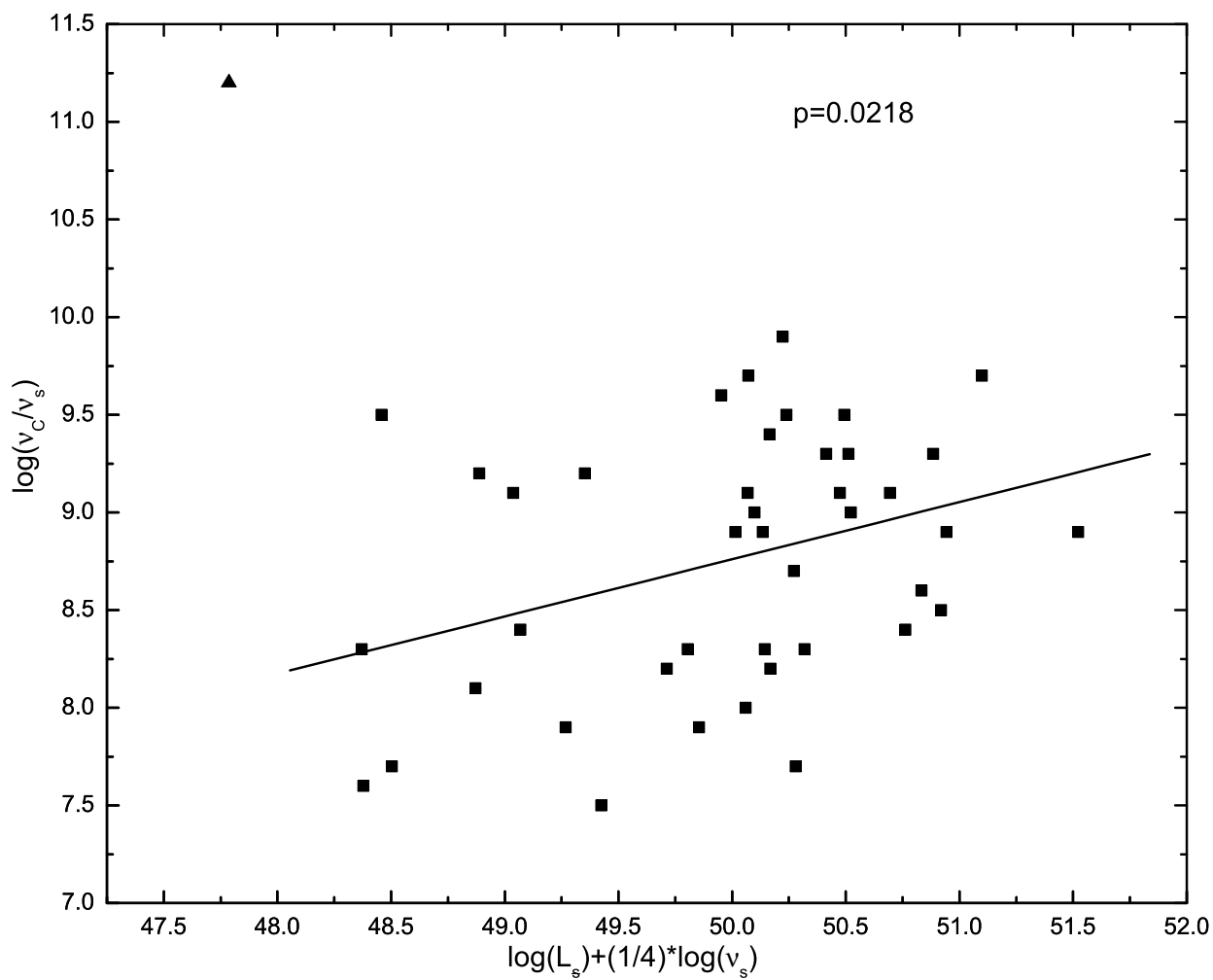}
\caption{The correlation between the parameter $L_{s}\nu_{s}^{1/4}$ and the ratio of Compton to synchrotron peak frequencies $r_{Cs}\equiv\nu_{C}/\nu_{s}$. The blazar 0FGL J1719.3+1746, as indicated by the triangle at upper left corner, shows extreme ratio $r_{Cs}$. The best fitting of the sample, excluding 0FGL J1719.3+1746, gives $p=0.0218$.}
\label{beaming_mass_r}
\end{figure}

\begin{figure}
\epsscale{1}
\plotone{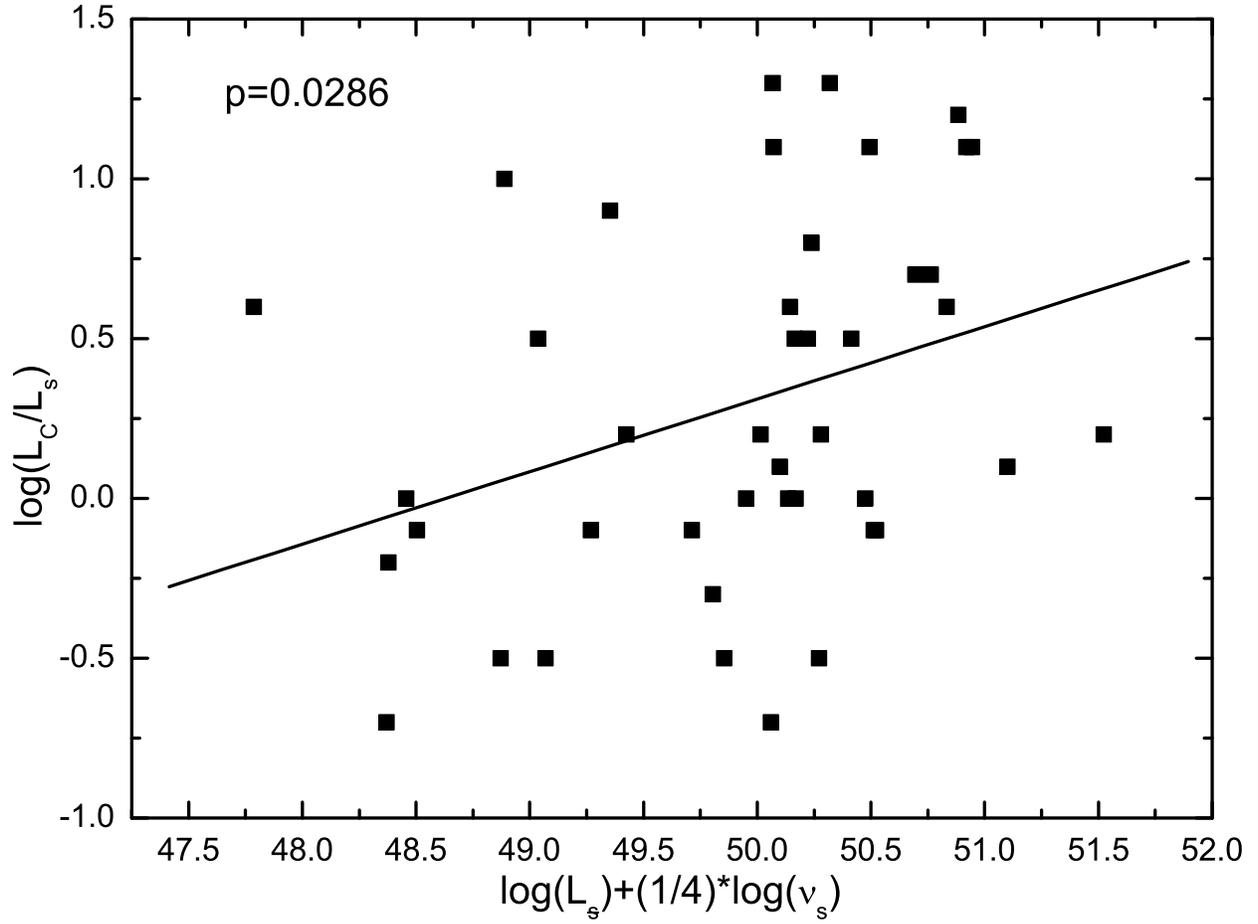}
\caption{The correlation between the parameter $L_{s}\nu_{s}^{1/4}$ and the ratio of Compton to synchrotron luminosities $CD\equiv L_{C}/L_{s}$. The best fitting gives $p=0.0286$.}
\label{beaming_mass_cd}
\end{figure}

\begin{figure}
\epsscale{1}
\plotone{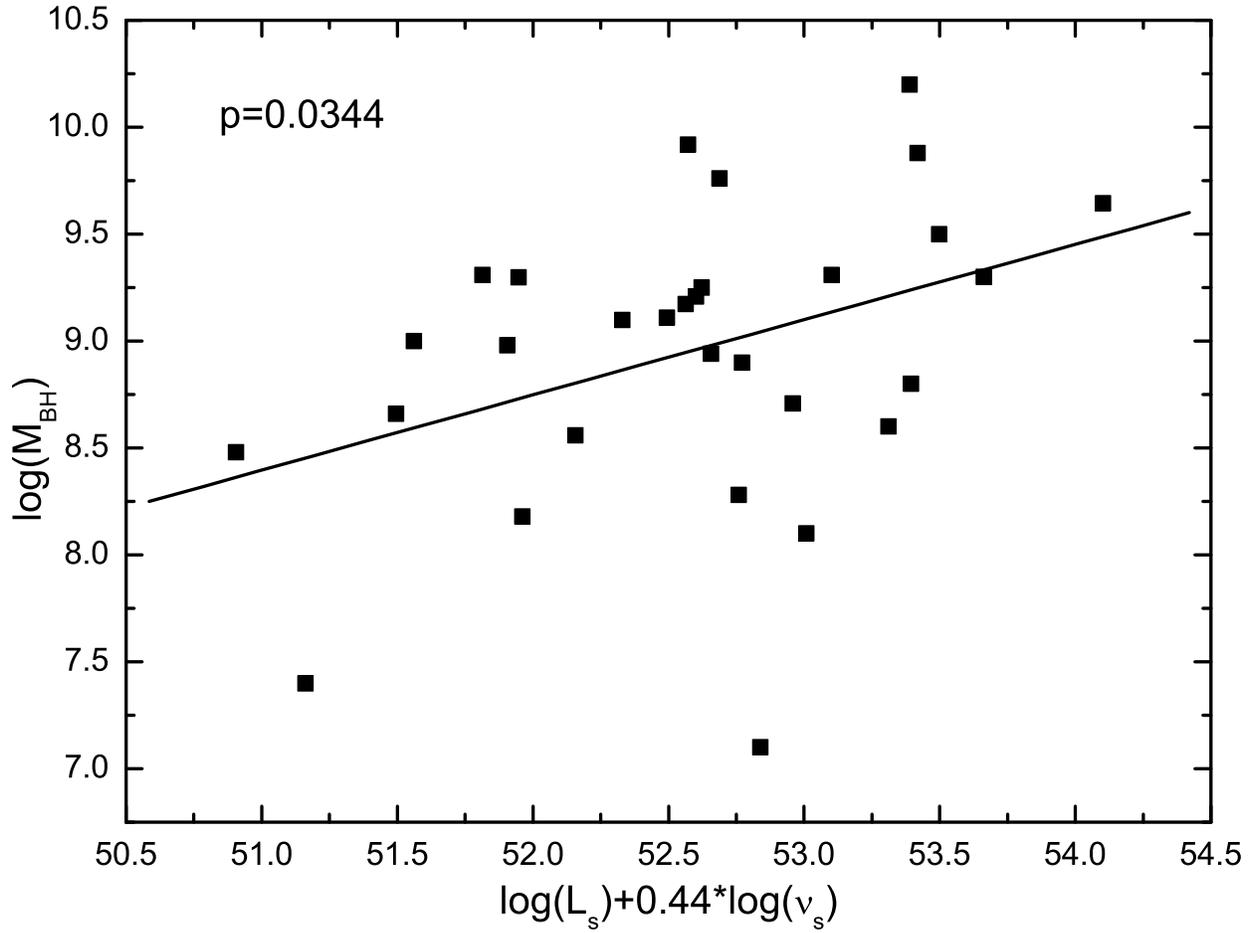}
\caption{The correlation between the parameter $L_{s}\nu_{s}^{0.44}$ and the black hole masses. The best fitting indicates $p=0.0344$.}
\label{low_mass}
\end{figure}

\begin{figure}
\epsscale{1}
\plotone{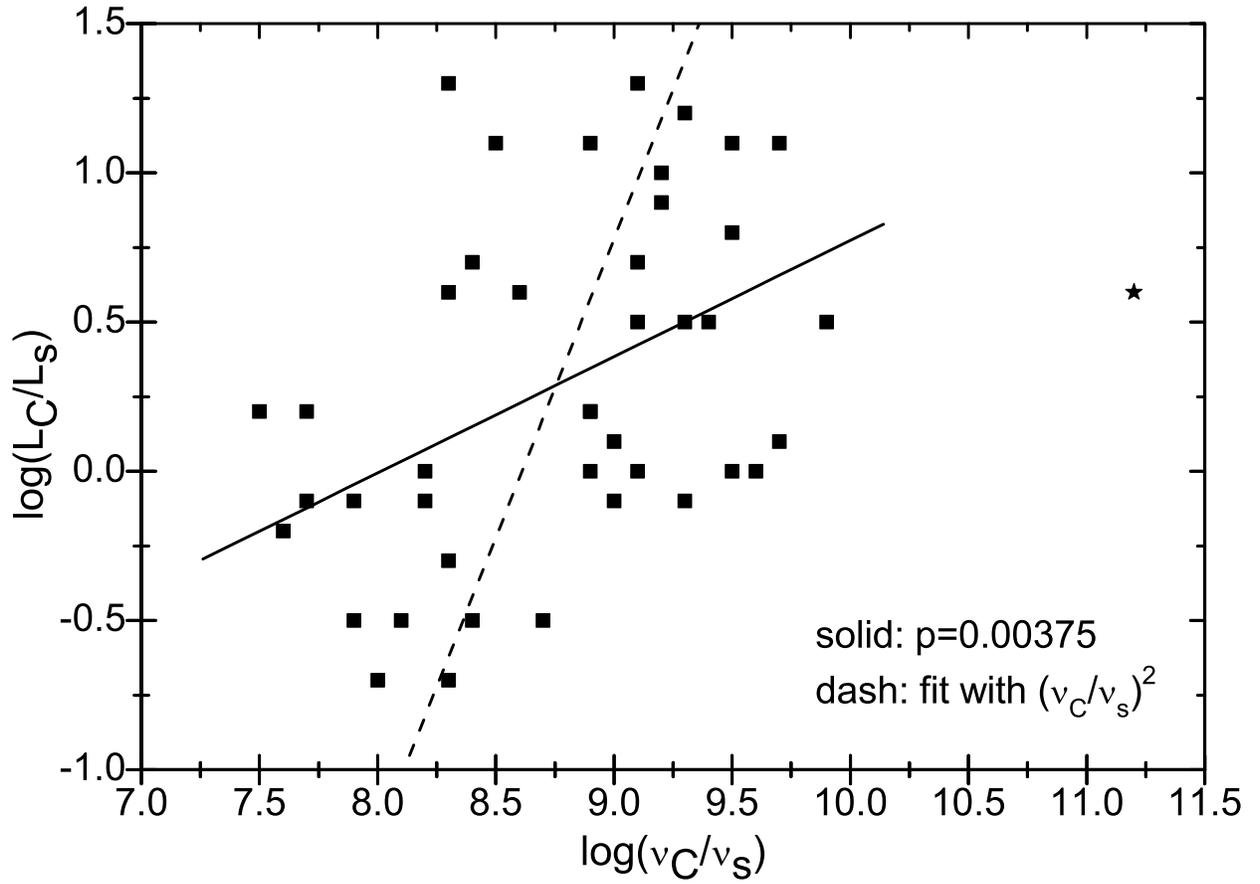}
\caption{The correlation between $r_{Cs}$ and $CD$. The solid line present the best fitting ($p=0.00375$). The dashed line is the best fitting with relation $L_{IC}/L_{sy}\propto\left(\nu_{IC}^{p}/\nu_{sy}^{p}\right)^{2}$.}
\label{EC3}
\end{figure}

\begin{figure}
\epsscale{1}
\plotone{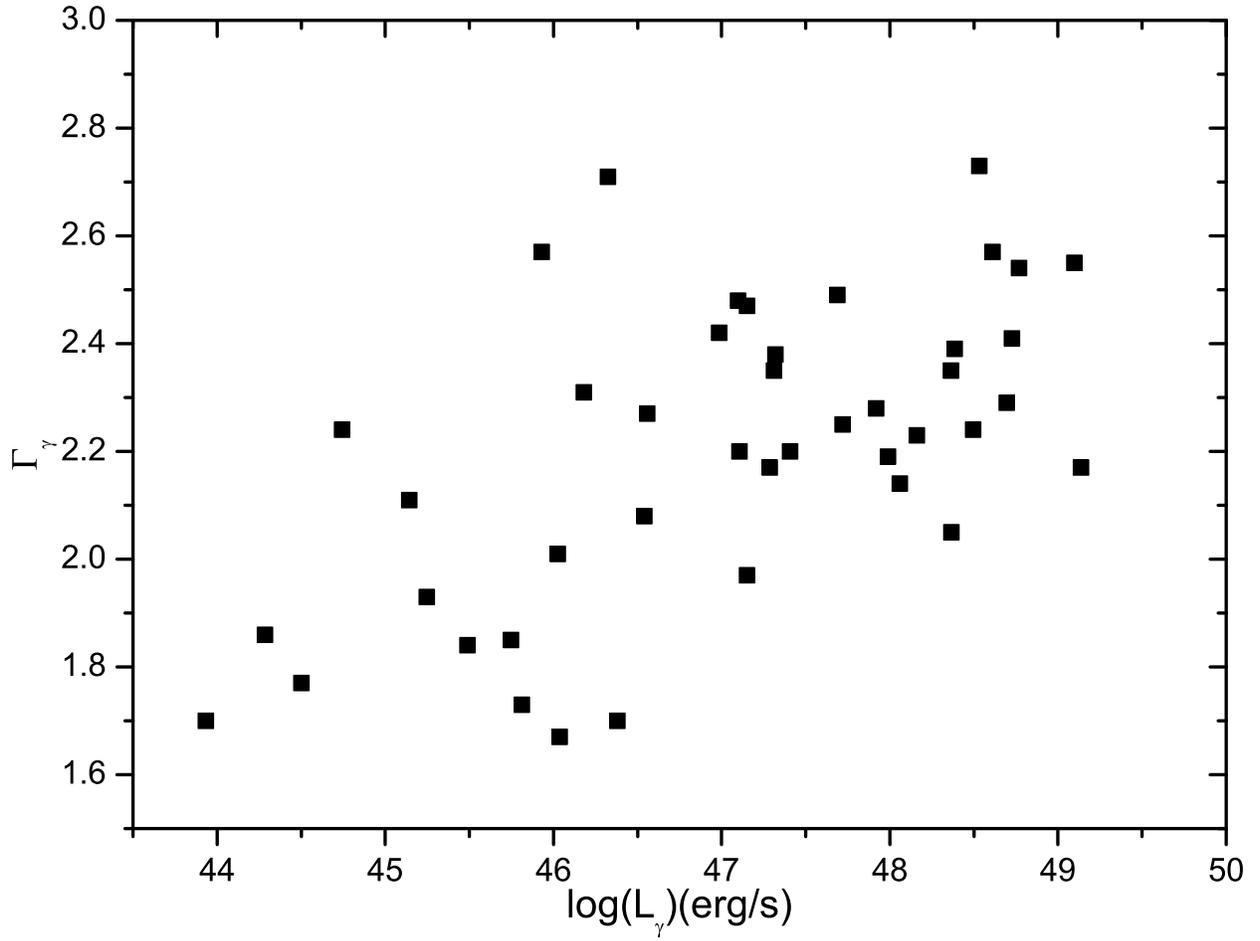}
\caption{$\gamma$-ray luminosity ($L_{\gamma}$) vs. $\gamma$-ray photon indexes ($\Gamma_{\gamma}$) with $p=2.71\times10^{-5}$.}
\label{lum_gamma}
\end{figure}

\begin{figure}
\epsscale{1}
\plotone{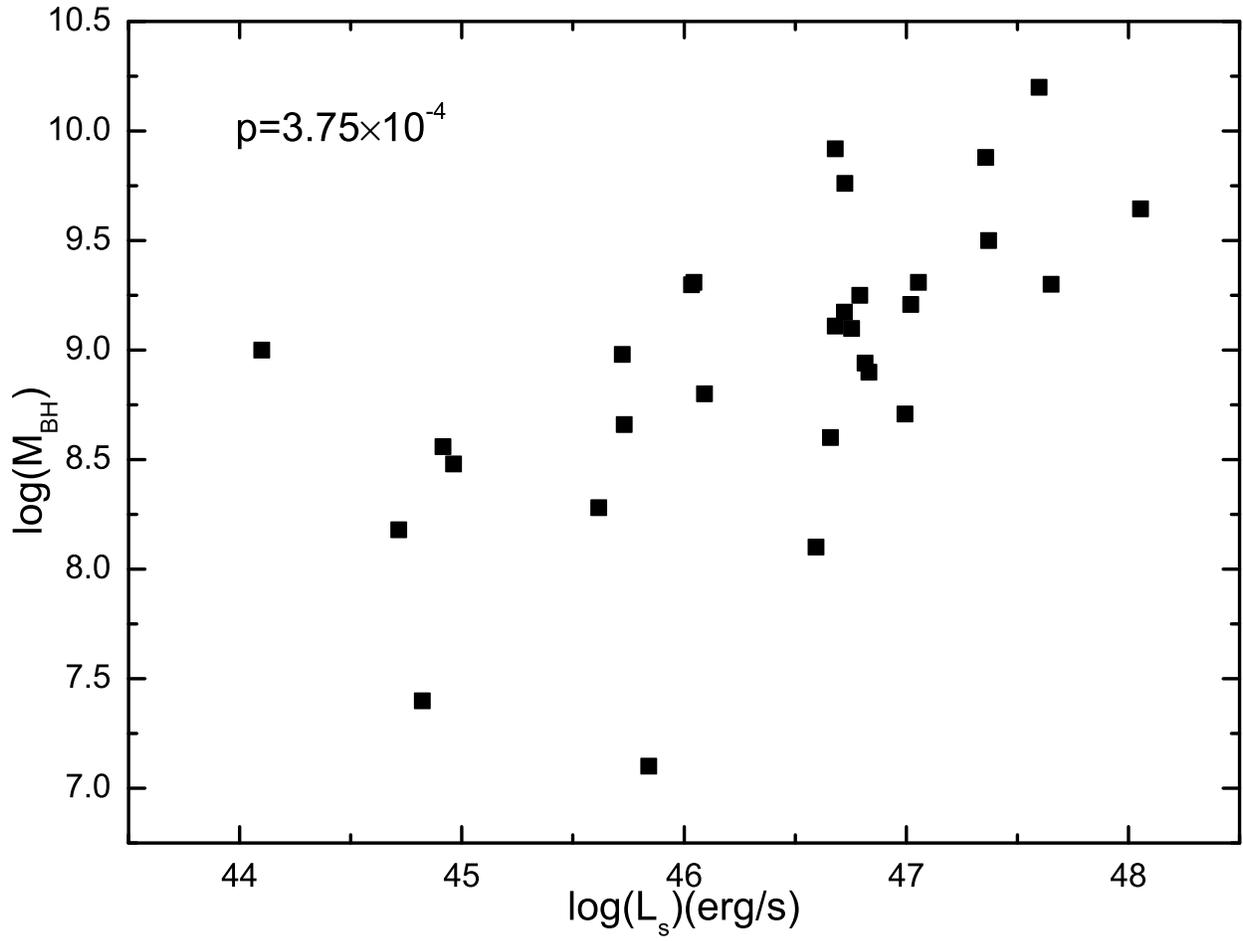}
\caption{Synchrotron peak luminosity ($L_{s}$) vs. black hole mass with $p=3.75\times10^{-4}$.}
\label{lum_mass}
\end{figure}

\begin{figure}
\epsscale{1}
\plotone{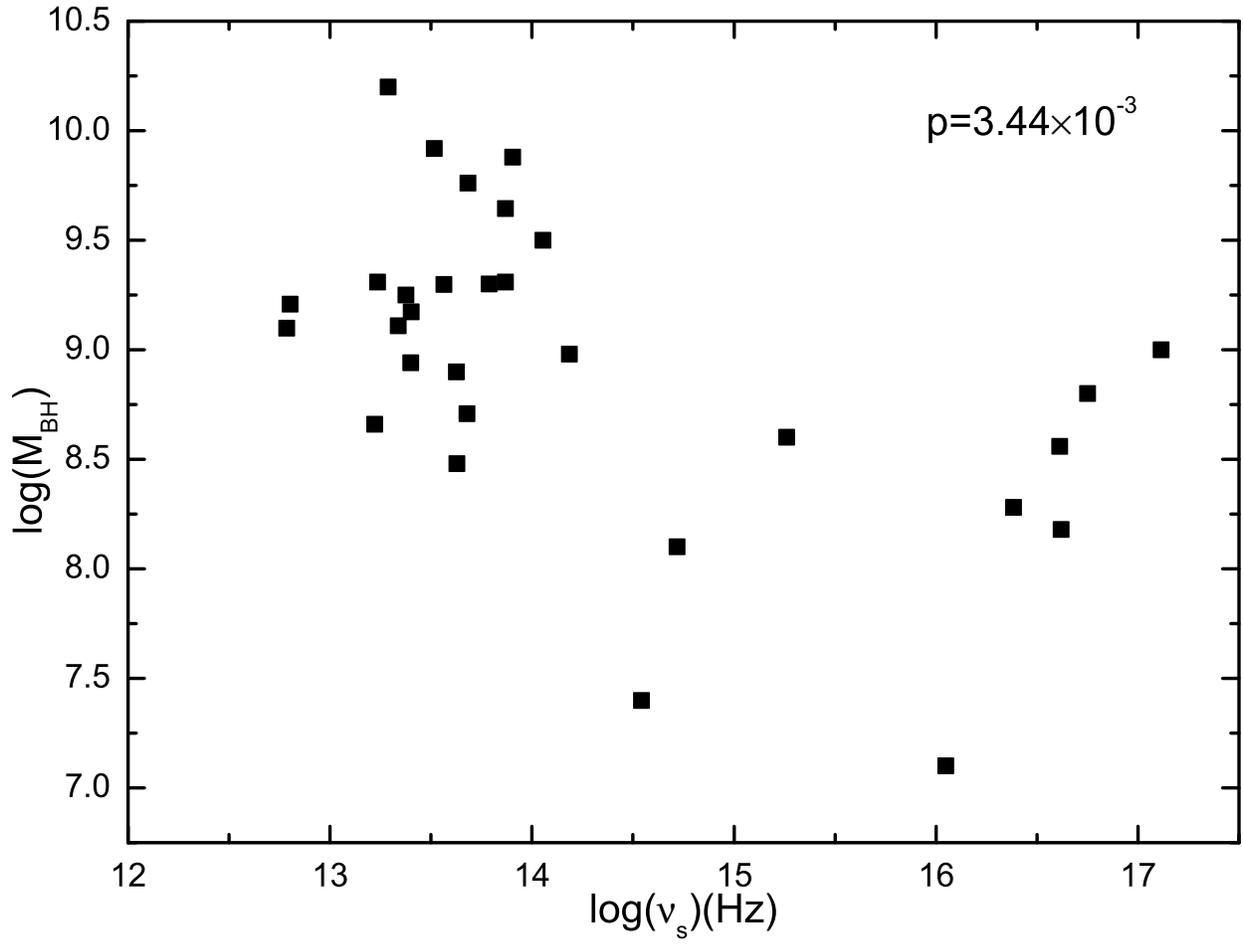}
\caption{Synchrotron peak frequency ($\nu_{s}$) vs. black hole mass with $p=3.44\times10^{-3}$.}
\label{nu_mass}
\end{figure}

Our results suggest that it is not the beaming effect but the black hole mass accounting for the properties of the low $\nu_{s}$ low $L_{s}$ blazars. In drawing the conclusion, there are caveats should be noted. From figures \ref{beaming_mass_r} and \ref{beaming_mass_cd}, it can be seen that both correlations are not strict, but have large scattering. This means that the beaming effect also can play a certain role although not determines the nature of low $\nu_{s}$ low $L_{s}$ sources. Many radio galaxies are detected by \emph{Fermi}/LAT. Within unified model of radio loud AGN, radio galaxies are the parent population of blazars but with large viewing angle. Figure 24 in \citet{2010ApJ...715..429A} presented the correlation between $\gamma$-ray photon spectral index and $\gamma$-ray luminosity, including the radio galaxies. It can be seen that radio galaxies have lower luminosity and average softer spectra relative to blazars. This is qualitatively consistent with the hypothesis that misaligned sources have lower luminosity and lower peak frequency. Black hole masses of 30 of 43 blazars are collected. These blazars show significant correlation between luminosity $L_{s}$ and black hole mass ($p=3.75\times10^{-4}$, figure \ref{lum_mass}), and also present an inverse correlation between peak frequency $\nu_{s}$ and black hole mass ($p=3.44\times10^{-3}$, figure \ref{nu_mass}). This indicates that the high peak frequency blazars have lower black hole masses. Through the correlation between the black hole mass and $L_{s}\nu_{s}^{0.44}$, we showed that the low $\nu_{s}$ low $L_{s}$ balzars may have smaller black hole masses. The slope ($s=0.44$) is derived from the best fitting. As we showed, the $\log\nu_{s}-\log L_{s}$ plane is more like wedge-shape. The upper boundary of the wedge-shape seems steeper than $s=0.44$ (see figure \ref{nu_lum}). On the other hand, if we linearly fit the $\log\nu_{s}-\log L_{s}$ plane excluding the low $\nu_{s}$ low $L_{s}$ blazars, the fitting slope will be steeper than $s=0.44$. We then choose a steeper slope ($s=0.6$) and correlate the parameter $L_{s}\nu_{s}^{0.6}$ with the black hole mass. The result presents very poor correlation ($p=0.2$). Therefore, it seems that lower black hole mass can account for these low $\nu_{s}$ low $L_{s}$ blazars, but the nature can not be definitely determined. To check the results, larger sample are needed. 1LAC \citep{2010ApJ...715..429A} supplies a huge amount of data, which can help to determine the properties of IC component. The multi-band SEDs can be derived from ground and space observatories. The black hole masses can be derived using a uniform method. The information about quasi-simultaneous SEDs for the latter sample would probably be less complete than for our sample, but its richness will yield interesting results.

If those blazars are really having smaller black hole, this does not support the sequence $\nu_{s}-L_{s}$ inverse correlation, but it is still consistent with the sequence $\gamma_{b}-U_{tot}$ inverse correlation. Here, we call $\nu_{s}-L_{s}$ the phenomenological sequence and $\gamma_{b}-U_{tot}$ the theoretical sequence \citep[see][]{2008MNRAS.387.1669G}. As suggested by \citet{2008MNRAS.387.1669G}, blazars with smaller black hole can have jet energy dissipated within the BLR. Following the theoretical sequence, the high energy electron in jet will suffer larger cooling, and then smaller $\gamma_{b}$. This results in a lower synchrotron peak frequency and lower luminosity. So, our result can be regarded as departure from the phenomenal sequence, but consistent with the theoretical sequence. The $\gamma_{b}-U_{tot}$ relation has different slopes from different studies, range from $1/2$ to $1$ \citep[see][]{2008MNRAS.385..283C, 1998MNRAS.301..451G, 2002A&A...386..833G, 2009MNRAS.399.2041G, 2010MNRAS.402..497G}. The reason accounting for this relation is not clear. \citet{2002A&A...386..833G} suggest that $\gamma_{b}\propto U_{tot}'^{-1}$ implies a constant cooling time at peak frequency, which may correspond to a constant light crossing time. The relation $\gamma_{b}\propto U_{tot}'^{-1/2}$ may denote a constant heating rate \citep[see][]{1999AN....320..232G}.

The correlation between the ratio $r_{Cs}$ and $CD$ is a new result. These two parameters are independent of redshift or beaming effect. They may be related to the jet conditions and radiative processes. Within the leptonic model, the relation between IC and synchrotron components implicates the relative importance of EC to SSC, at least on statistics. Here we gave an explanation: it may be the result of EC dominant. This is consistent with the detailed SED modeling \citep[see][]{2009MNRAS.399.2041G, 2010MNRAS.402..497G}. Some blazars present long term outbursts. Given a blazar, the emission regions of different outburst/quiet states may be surrounded by similar external radiation field, e.g., BLR photons. In this case, the EC and synchrotron emissions will follow the equation \ref{eq_cd_rcs}. For some extreme blazars, e.g., 3C 279, if we have SEDs at different outburst/quiet states, these combining with equation 4 will yield interesting results. The caveat is that the equation is derived from one zone symmetrical model. Enlarging sample to check the above correlation is of course needed.

In summary, we presented the plane $\log\nu_{s}-\log L_{s}$ for bright \emph{Fermi} blazars. The plane shows inverse correlation statistically, but some low $\nu_{s}$ low $L_{s}$ blazars appear. These blazars may be characterized by relatively smaller black hole masses rather than by weaker beaming. 
The ratio $r_{Cs}$ correlates with the Compton dominance $CD$. This may indicate that in most blazars the high energy emission is dominated by the External Compton process.

\acknowledgments

We thank the anonymous referee for insightful comments and constructive suggestions. We are grateful to Xinwu Cao, Yi Liu, Hongtao Liu and Fan Li for helpful discussions. We thank supports of the National Natural Science Foundation of China (Grant Nos. 10903025, 10778702, 10973034 and 10833002) and the 973 Program (Grant No. 2009CB824800).


\begin{deluxetable}{lrrrrrrrrrr}
\tabletypesize{\footnotesize}
\tablecaption{Data of Selected 43 Blazars}
\tablehead{
\colhead{Name(0FGL)}&
\colhead{$\log\nu_{s}$}&
\colhead{$\log L_{s}$}&
\colhead{$\log\nu_{C}$}&
\colhead{$\log L_{C}$}&
\colhead{$z$}&
\colhead{$\Gamma_{\gamma}$}&
\colhead{$\log L_{\gamma}$}&
\colhead{Type}\tablenotemark{a}&
\colhead{$\log M_{BH}$}&
\colhead{ref.}\tablenotemark{b}\\
\colhead{(1)}&
\colhead{(2)}&
\colhead{(3)}&
\colhead{(4)}&
\colhead{(5)}&
\colhead{(6)}&
\colhead{(7)}&
\colhead{(8)}&
\colhead{(9)}&
\colhead{(10)}&
\colhead{(11)}
}
\startdata
J0033.6-1921  &    16.3  &    46.1  &    24.5  &    46.1  &   0.610  &   1.70  &   46.4  &   BL    &           &       \\
J0137.1+4751  &    13.9  &    47.1  &    22.9  &    47.0  &   0.859  &   2.20  &   47.4  &   FSRQ  &    9.309  &   c02 \\
J0210.8-5100  &    12.8  &    47.0  &    22.7  &    47.5  &   1.003  &   2.28  &   47.9  &   FSRQ  &    9.208  &   f04 \\
J0222.6+4302  &    15.3  &    46.7  &    24.4  &    46.7  &   0.444  &   1.97  &   47.2  &   BL    &    8.600  &   l03 \\
J0229.5-3640  &    14.0  &    46.8  &    22.3  &    48.1  &   2.115  &   2.57  &   48.6  &   FSRQ  &           &       \\
J0238.4+2855  &    13.1  &    47.2  &    22.4  &    47.1  &   1.213  &   2.49  &   47.7  &   FSRQ  &           &       \\
J0238.6+1636  &    13.8  &    47.7  &    23.5  &    47.8  &   0.940  &   2.05  &   48.4  &   BL    &    9.300  &   l03 \\
J0349.8-2102  &    13.5  &    47.6  &    22.4  &    48.7  &   2.944  &   2.55  &   49.1  &   FSRQ  &           &       \\
J0423.1-0112  &    13.7  &    46.7  &    22.0  &    47.3  &   0.915  &   2.38  &   47.3  &   FSRQ  &    9.760  &   c02 \\
J0428.7-3755  &    13.6  &    46.8  &    23.1  &    47.6  &   1.112  &   2.14  &   48.1  &   BL    &    8.900  &   d10 \\
J0449.7-4348  &    15.7  &    45.9  &    24.0  &    45.6  &   0.205  &   2.01  &   46.0  &   BL    &           &       \\
J0457.1-2325  &    13.4  &    46.7  &    23.1  &    47.8  &   1.003  &   2.23  &   48.2  &   FSRQ  &    9.173  &   f04 \\
J0507.9+6739  &    16.8  &    46.1  &    24.5  &    46.3  &   0.416  &   1.67  &   46.0  &   BL    &    8.800  &   f03 \\
J0531.0+1331  &    13.3  &    47.6  &    21.8  &    48.7  &   2.070  &   2.54  &   48.8  &   FSRQ  &   10.200  &   l03 \\
J0538.8-4403  &    13.7  &    47.0  &    23.0  &    47.5  &   0.892  &   2.19  &   48.0  &   BL    &    8.709  &   f04 \\
J0722.0+7120  &    14.7  &    46.6  &    23.4  &    46.1  &   0.310  &   2.08  &   46.5  &   BL    &    8.100  &   l03 \\
J0730.4-1142  &    13.5  &    47.1  &    23.0  &    48.2  &   1.589  &   2.29  &   48.7  &   FSRQ  &           &       \\
J0855.4+2009  &    13.5  &    46.7  &    21.5  &    46.0  &   0.306  &   2.31  &   46.2  &   BL    &    9.919  &   f04 \\
J0921.2+4437  &    13.9  &    47.4  &    22.5  &    48.0  &   2.190  &   2.35  &   48.4  &   FSRQ  &    9.880  &   c09 \\
J1015.2+4927  &    16.4  &    45.6  &    24.6  &    45.5  &   0.212  &   1.73  &   45.8  &   BL    &    8.280  &   w08 \\
J1058.9+5629  &    14.7  &    44.8  &    22.4  &    44.7  &   0.143  &   2.11  &   45.1  &   BL    &           &       \\
J1057.8+0138  &    13.4  &    46.8  &    22.3  &    46.8  &   0.888  &   2.20  &   47.1  &   BZU   &    9.250  &   c09 \\
J1104.5+3811  &    16.6  &    44.9  &    25.0  &    44.4  &   0.030  &   1.77  &   44.5  &   BL    &    8.560  &   w08 \\
J1159.2+2912  &    13.3  &    46.7  &    22.2  &    46.9  &   0.729  &   2.47  &   47.2  &   FSRQ  &    9.110  &   c09 \\
J1221.7+2814  &    14.5  &    44.8  &    24.0  &    44.8  &   0.102  &   1.93  &   45.2  &   BL    &    7.400  &   l03 \\
J1229.1+0202  &    13.6  &    46.0  &    21.1  &    46.2  &   0.158  &   2.71  &   46.3  &   FSRQ  &    9.298  &   f04 \\
J1256.1-0548  &    12.8  &    46.8  &    22.4  &    46.8  &   0.536  &   2.35  &   47.3  &   FSRQ  &    9.099  &   c02 \\
J1310.6+3220  &    13.4  &    46.8  &    22.8  &    47.3  &   0.997  &   2.25  &   47.7  &   FSRQ  &    8.940  &   c09 \\
J1457.6-3538  &    14.0  &    47.2  &    23.1  &    47.9  &   1.424  &   2.24  &   48.5  &   FSRQ  &           &       \\
J1504.4+1030  &    14.1  &    47.4  &    23.4  &    48.6  &   1.839  &   2.17  &   49.1  &   FSRQ  &    9.500  &   c09 \\
J1512.7-0905  &    13.2  &    46.0  &    22.4  &    46.9  &   0.360  &   2.48  &   47.1  &   FSRQ  &    9.310  &   c02 \\
J1522.2+3143  &    13.7  &    46.6  &    22.8  &    47.9  &   1.487  &   2.39  &   48.4  &   FSRQ  &           &       \\
J1653.9+3946  &    17.1  &    44.1  &    24.7  &    43.9  &   0.033  &   1.70  &   43.9  &   BL    &    9.000  &   w08 \\
J1719.3+1746  &    13.6  &    44.4  &    24.8  &    45.0  &   0.137  &   1.84  &   45.5  &   BL    &           &       \\
J1751.5+0935  &    13.2  &    45.7  &    22.3  &    46.2  &   0.322  &   2.27  &   46.6  &   BL    &    8.660  &   f03 \\
J1849.4+6706  &    13.7  &    46.7  &    22.7  &    46.8  &   0.657  &   2.17  &   47.3  &   FSRQ  &           &       \\
J2000.2+6506  &    16.6  &    44.7  &    24.7  &    44.2  &   0.047  &   1.86  &   44.3  &   BL    &    8.180  &   w08 \\
J2143.2+1741  &    14.2  &    45.7  &    22.1  &    45.6  &   0.213  &   2.57  &   45.9  &   FSRQ  &    8.980  &   f03 \\
J2158.8-3014  &    16.0  &    45.8  &    23.9  &    45.3  &   0.116  &   1.85  &   45.7  &   BL    &    7.100  &   l03 \\
J2202.4+4217  &    13.6  &    45.0  &    21.9  &    44.3  &   0.069  &   2.24  &   44.7  &   BL    &    8.480  &   w08 \\
J2254.0+1609  &    13.9  &    48.1  &    22.8  &    48.3  &   0.859  &   2.41  &   48.7  &   FSRQ  &    9.644  &   f04 \\
J2327.3+0947  &    13.6  &    47.4  &    22.0  &    48.1  &   1.843  &   2.73  &   48.5  &   FSRQ  &           &       \\
J2345.5-1559  &    13.5  &    45.5  &    22.7  &    46.5  &   0.621  &   2.42  &   47.0  &   FSRQ  &           &       \\
\enddata
\tablenotetext{a}{BL is the abbreviation of BL Lac; BZU denotes blazar of unknown type \citep[see][]{2009arXiv0912.2040A}.}
\tablenotetext{b}{References, C02: \citet{2002MNRAS.331..111C}; C09: \citet{2009RAA.....9.1192C}; D10: \citet{2010arXiv1011.5879D}; F03: \citet{2003MNRAS.343..505F}; F04: \citet{2004ApJ...602..103F}; L03: \citet{2003MNRAS.340..632L}; W08: \citet{2008MNRAS.385..119W}.}
\tablecomments{Column (1) provides the LAT name of the source. (2) and (3) indicate the synchrotron peak frequency and luminosity. (4) and (5) denote the IC peak frequency and luminosity. The redshift, $\gamma$-ray photon indices $\Gamma_{\gamma}$, $\gamma$-ray luminosity $L_{\gamma}$ and the optical classification are listed in Columns (6), (7), (8) and (9), respectively. Columns (10) and (11) are the black hole masses and the references. Data from \citet{2009ApJ...700..597A, 2009arXiv0912.2040A}.}
\label{blazar}
\end{deluxetable}

\begin{deluxetable}{lrrr}
\tabletypesize{\footnotesize}
\tablecaption{Data of four \emph{Fermi} detected NLS1s}
\tablehead{
\colhead{Name}&
\colhead{$\log\nu_{s}$}&
\colhead{$\log L_{s}$}\\
\colhead{(1)}&
\colhead{(2)}&
\colhead{(3)}
}
\startdata
1H 0323+342     &    13.75  &    44.39  &  \\
PMN J0948+0022  &    12.94  &    45.43  &  \\
PKS 1502+036    &    13.02  &    45.14  &  \\
PKS 2004-447    &    13.05  &    44.55  &  \\
\enddata
\tablecomments{Column (1) provides the name of the source. We use quadratic polynomial to fit the SED of the low component emissions of the four NLS1s and get the synchrotron peak frequency and luminosity, which are listed in columns (2) and (3). \citep[See][]{2009ApJ...707L.142A}}
\label{NLS1}
\end{deluxetable}

\end{document}